\documentclass[aps,amsmath,amssymb,reprint,superscriptaddress,prd]{revtex4-2}

\usepackage{txfonts}
\usepackage{graphicx}
\usepackage{subfigure}
\usepackage{bm}
\usepackage{dcolumn}
\usepackage{braket}
\usepackage{color}
\usepackage[colorlinks,linkcolor=red,anchorcolor=blue,citecolor=blue]{hyperref}

\begin{document}


\title{Dynamical instability and transport peak of chiral matter from holography}


\author{Pei Zheng}
\email{zhengp@stu.pku.edu.cn}
\affiliation{School of Nuclear Science and Technology, University of Chinese Academy of Sciences, Beijing 100049, China}
\affiliation{Department of Physics and State Key Laboratory of Nuclear Physics and Technology, Peking University, Beijing 100871, China}

\author{Yidian Chen}
\email{chenyidian@hznu.edu.cn}
\affiliation{School of Physics, Hangzhou Normal University, Hangzhou, 311121, China}

\author{Danning Li}
\email{lidanning@jnu.edu.cn}
\affiliation{Department of Physics and Siyuan Laboratory, Jinan University, Guangzhou 510632, China}

\author{Mei Huang}
\email{huangmei@ucas.ac.cn}
\affiliation{School of Nuclear Science and Technology, University of Chinese Academy of Sciences, Beijing 100049, China}

\author{Yu-xin Liu}
\email{yxliu@pku.edu.cn}
\affiliation{Department of Physics and State Key Laboratory of Nuclear Physics and Technology, Peking University, Beijing 100871, China}
\affiliation{Center for High Energy Physics, Peking University, Beijing 100871, China}

\begin{abstract}
We study dynamical properties of strongly coupled chiral matter by using holographic method. We demonstrate, at both linear and nonlinear levels, that perturbations on thermodynamically unstable backgrounds within the spinodal region of chiral first-order phase transitions exhibit dynamic instability. The corresponding magnitude of dynamic instability can be characterized by the critical momentum. Furthermore, we found that, within a certain temperature range, the quasi-normal mode spectrum contains purely imaginary diffusive modes. As spatial momentum increases, a transition occurs in the system's long-time dynamics. The dominant contribution shifts from diffusive mode to propagating mode. When the diffusive mode becomes dominant, the spectral function exhibits a transport peak structure in the low-frequency region. A heuristic argument suggests that this particular transition can be related to the chiral symmetry breaking and restoration. 
\end{abstract}

\maketitle

\section{Introduction\label{sec:intro}}

Research on the properties of strong interaction matter has consistently remained a highly active field in theoretical physics. Due to the strong-coupling nature of quantum chromodynamics (QCD) describing strong interactions in the low-energy region, commonly employed perturbative approaches become inapplicable. Since issues of primary concern, such as spontaneous breaking of chiral symmetry and color confinement, are intrinsically linked to low-energy scale physics, in-depth investigation of these phenomena necessitates the development of appropriate nonperturbative field theoretic approaches. Although some methods based on functional integral have made significant progress in studying different aspects of QCD matter \cite{Fischer:2018sdj,Dupuis:2020fhh}, current research on the dynamical behavior of strongly coupled matter remains insufficient in both systematic rigor and comprehensive scope, as existing functional methods encounter certain difficulties in addressing dynamical issues. 

Holographic AdS/CFT correspondence \cite{Witten:1998qj,Witten:1998zw,Maldacena:1997re,Gubser:1998bc} (see \cite{Aharony:1999ti} for a review) provides a suitable tool to study strongly coupled systems. This correspondence asserts that string theories lived on (asymptotically) anti-de Sitter spacetime are dual to some quantum field theories in lower dimension. In the low-energy limit, the string theories can be described by appropriate classical supergravity theories. So the problems in quantum field theories can be mapped to problems in classical gravity theories. One of the incredible features is that such duality is the so-called strong/weak duality. To illustrate, physical information in strongly coupled field theories is encoded in weakly coupled gravitational theories by using such duality. This implies that non-perturbative features of strongly coupled field theoretic systems can be extracted by directly studying weakly coupled gravitational theories. 

Based on AdS/CFT correspondence, small perturbations of gravity background are interpreted as small perturbations in the dual field theory. This particular situation in holographic theory can be made more precise by checking spectrum of the so-called quasi-normal modes of (asymptotically) AdS spacetime. Typically, quasi-normal modes are solutions to linearized equation of motions satisfied by all kinds of fluctuations \cite{Berti:2009kk,Kovtun:2005ev,Starinets:2002br,Horowitz:1999jd}. In particular, the boundary conditions should be chosen specifically. Considering the fact that the future horizon do not emit radiation classically, one choose the incoming waves at the future horizon only. Consequently, the associated boundary value problem becomes non-Hermitian and the corresponding eigenvalues are complex. From the dual field theory perspective, these quasi-normal modes can be directly related to the poles of retarded Green's functions \cite{Son:2002sd}.    

From the gravity perspective, Gregory and Laflamme has already found that there exists a classical instability of higher dimensional black string/brane solutions, which is usually referred to the Gregory-Laflamme (GL) instability \cite{Gregory:1993vy,Gregory:1994bj}. In the AdS/CFT correspondence, GL instability rooted in classical gravitational frameworks has acquired novel physical significance. A natural interpretation is that the dual field theory exhibits corresponding dynamical instability. Typically, there is a thermodynamic instability in the field theory at finite temperature, like the spinodal region of a first-order phase transition. Gubser and Mitra naturally conjectured that the system with locally thermodynamical instability will induce dynamical instability \cite{Gubser:2000ec,Gubser:2000mm}. The conjecture linking thermodynamic and dynamical properties has been validated in subsequent studies across diverse theoretical frameworks \cite{Buchel:2005nt,Hirayama:2002hn,Miyamoto:2007mh,Reall:2001ag}. Furthermore, from an alternative perspective grounded in phenomenological consideration, potential dynamical instabilities in strongly coupled chiral QCD matter can also be explored based on the aforementioned conjecture. Regarding the importance of dynamical instabilities manifested in QCD matter, a case in point is that the instabilities of quark-gluon plasma (QGP) are believed to play an important role in the early stages of QGP equilibration \cite{Moore:2005rp,Arnold:2005ef,Arnold:2005qs,Kurkela:2011ub}. Besides, the so-called spatially modulated phase \cite{Nakamura:2009tf,Ooguri:2010xs} is found to be indicated by some dynamical instability of the typical homogeneous QGP phase \cite{Ooguri:2010kt,Demircik:2024aig,CruzRojas:2024igr}. Owing to all these considerations, the fundamental motivation of this work can be summarized as investigating the potential dynamical instability of QCD chiral matter and meanwhile, examining the validity of the Gubser-Mitra conjecture within this specific physical context. 

The paper is organized as follows. In Sec. \ref{sec:model}, we briefly introduce the model used in this model and related chiral phase transition behaviors. In Sec. \ref{sec:dynamical instability}, we study the dynamical properties of chiral system in the spinodal region of first-order chiral phase transition either in linear perturbation level or in full non-linear evolution level. Some interesting transition behavior is also observed while studying the quasi-normal spectrum. In Sec. \ref{sec:diffusion-to-sound}, we investigate such diffusion-to-sound transition by calculating quasi-normal modes and also corresponding spectral functions. In Sec. \ref{sec:sum}, we provide a summary and further discussion.  

\section{Basic holographic model setup and chiral phase transition\label{sec:model}}

Within the holographic framework, the hard-wall and soft-wall AdS/QCD models were first proposed in \cite{Erlich:2005qh,Karch:2006pv}.  From bottom-up perspective, the 4D global symmetry $SU(N_f)_L\times SU(N_f)_R$ of chiral system is directly promoted to 5D gauge symmetry of the corresponding bulk system. By incorporating desired symmetry structure, these effective models are suitable to study problems related to the spontaneous symmetry breaking of chiral symmetry. Based on the hard-wall model, a quadratic dilaton model was introduced in the soft-wall model to describe the linear trajectory of the higher excitations. At finite temperature, after introducing an additional scale relevant for chiral dynamics, the phase diagram obtained from the soft-wall model \cite{Chelabi:2015gpc,Li:2016smq,Chen:2018msc,Ballon-Bayona:2024twa,Bartz:2024dgd,Fang:2018vkp,Rodrigues:2018pep,Ahmed:2024rbj} was shown to qualitatively agree with the so-called `Columbia plot' obtained by combining lattice simulations and other effective methods. In addition to the order parameter, the thermal properties of hadrons, in particular of the Goldstone bosons,  were shown to qualitatively  agree \cite{Cao:2022csq,Cao:2021tcr,Cao:2020ryx} with the four-dimensional finite temperature chiral perturbation theory \cite{Son:2001ff}. Besides the equilibrium studies, it is also very convenient to extend this model to study the nonequilibrium phase transition. Near the second order transition, a non-trivial dynamic phenomena called `prethermalization' was shown to appear in the intermediate time \cite{Cao:2022mep,Zheng:2024rzl,Giannuzzi:2025fsv}. But the dynamic instabilities close to the first order phase transition was rarely studied within this model. Thus, it is still interesting to study the dynamic properties of the first-order phase transition in this model.

\subsection{The nonlinear soft-wall model}
\label{subsec:soft-wall model}

In this work, we use the conventions of \cite{Karch:2006pv} and write the 5D action of the nonlinear soft-wall model as
\begin{equation}
\label{eq:5D action}
    S=-\int d^5x\sqrt{-g}e^{-\Phi}{\rm{Tr}}\left[|D_mX|^2+V_X(X)+\dfrac{1}{4g^2_5}(F^2_L+F^2_R)\right],
\end{equation}
where
\begin{align}
\label{eq:5D convention}
    F^{L/R}_{mn}=&\partial_{m}A^{L/R}_n-\partial_n A^{(L/R)}_m-i \left[A^{L/R}_m,A^{L/R}_n\right], \nonumber \\
    D_mX=&\partial_mX-iA^L_m X+iXA^R_m.
\end{align}

In the original version of the soft-wall model, the dilaton field $\Phi$ is chosen to be a simple quadratic form $\Phi(z)=\mu^2 z^2$. By using this parametrization, the calculated meson spectra are consistent with desired linear Regge behavior \cite{Collins:1977jy}, i.e., $m^2_n\sim n$. However, it can be shown that there is no spontaneous chiral symmetry breaking in this case \cite{Chelabi:2015gpc}. So, this simple quadratic form of dilaton field can not fulfill our needs. 

Note that only the IR behavior of dilaton field can be constrained by the requirement of linear Regge trajectory. Consequently, there remains some freedom to modify the concrete behavior of dilaton field in the UV region. As pointed out in \cite{Chelabi:2015gpc}, the specific profile of dilaton field which serves our purpose well can be written as 
\begin{equation}
\label{eq:dilaton profile}
    \Phi(z)= -\mu^2_1z^2+(\mu^2_1+\mu^2_0) z^2\tanh(\mu^2_2z^2).
\end{equation}

The nonlinear scalar potential in this model is chosen to be
\begin{equation}
\label{eq:scalar potential}
    V_X(X)=M^2_5X^+X+\lambda|X|^4+\gamma{\rm{Re}}[\det(X)],
\end{equation}
with the third term on the right corresponding to the t 'Hooft determinant term. This term is necessary in $SU(3)$ or three flavor case. By adding this term, the chiral phase transition in the chiral limit is shown to be first order \cite{Li:2016smq}. Taking the AdS radius $L=1$ throughout this work, the 5D mass can be determined by standard AdS/CFT prescription, i.e., $M^2_5=(\Delta-p)(\Delta+p-4)$. For the complex scalar field $X$, the 5D mass square is just $M^2_5=-3$ by taking $\Delta=3$, $p=0$.  

There are five model parameters in this setup. As in \cite{Chelabi:2015gpc}, the two parameters in the scalar potential are taken to be
\begin{equation}
\label{eq:potential parameter}
    v_3\equiv\dfrac{2\sqrt{2}}{3} \gamma=-3,\ \ \ v_4\equiv\dfrac{\lambda}{4}=8,
\end{equation}
which are determined by the qualitative behavior of chiral phase transition in this model. The remaining three parameters in the dilaton profile \eqref{eq:dilaton profile} are taken to be
\begin{equation}
    \label{eq:dilaton parameters}
    \mu_0=0.43\text{GeV}, \ \ \ \mu_1=0.83\text{GeV}, \ \ \ \mu_2=0.176\text{GeV}.
\end{equation}
Here, the value of $\mu_0$ is determined by the slope of the radial excitation of mesons, i.e., $m^2_n\propto 4\mu^2_0 n$. The value of $\mu_1$ and $\mu_2$ are fixed by the chiral phase transition temperature and the value of chiral condensate at zero temperature for two flavor case. For more details, please refer to \cite{Chelabi:2015gpc}. 

Based on these values of model parameters, we will show the detailed chiral phase transition behavior from this model in the next section.

\subsection{Chiral phase transition}
\label{subsec:chiral PT}

By using the prescription proposed by Witten \cite{Witten:1998zw}, the gravity dual of finite temperature system corresponds to introducing a black hole in $AdS_5$.

In the probe limit, we neglect the back-reaction of matter fields to the background metric. Thus, the background metric is simply taken to be the AdS-Schwarzchild black hole solution in the Poincar\'{e} patch,
\begin{equation}
    \label{eq:AdS-SW}
    ds^2=\dfrac{1}{z^2} \left[-f(z) dt^2+\dfrac{1}{f(z)} dz^2+dx_idx^i\right],
\end{equation}
and
\begin{equation}
    \label{eq:blackening factor}
    f(z)=1-\dfrac{z^4}{z^4_h},
\end{equation}
where $z_h$ is the black hole horizon defined by $f(z_h)=0$. This parameter can be related to the temperature of boundary field theory directly by the following relation
\begin{equation}
    \label{eq:temperature}
    T=\Bigg|\dfrac{f^{\prime}(z_h)}{4\pi}\Bigg|=\dfrac{1}{\pi z_h}.
\end{equation}

In this work, we do not aim to study the detailed mechanisms of the chiral phase transition itself. Instead, we focus on the dynamical properties of the chiral system with non-trivial background chiral condensate. Hence, for simplicity, we choose the case with degenerate three-flavors ($m_q=m_u=m_d=m_s$) and the matrix field $X$ can be naturally parametrized as $X=\chi(z) \mathbf{1}_{3\times3}/\sqrt{2}$, where $1/\sqrt{2}$ is just a normalization factor in order to make kinetic term canonical.

By substituting the above metric ansatz \eqref{eq:AdS-SW} into the 5D action \eqref{eq:5D action}, the equation of motion of field $\chi(z)$ can be obtained by using standard variational method, which is
\begin{equation}
    \label{eq:background EOM}
    \chi^{\prime\prime}+\left(-\dfrac{3}{z}-\Phi^{\prime}+\dfrac{f^{\prime}}{f}\right)\chi^{\prime}+\dfrac{1}{z^2 f}(3\chi-3v_3\chi^2-4v_4\chi^3)=0,
\end{equation}
where the prime refers to the derivative with respect to AdS coordinate $z$. 

In order to extract the physical information, we need the asymptotic expansion of field $\chi(z)$ near the boundary ($z=0$). With the equation of motion \eqref{eq:background EOM}, this can be easily dealt with, and the result is written as 
\begin{align}
\label{eq:chi expansion}
\chi(z\rightarrow0)=&m_q \zeta z-3 m_q^2 v_3 \zeta^2z^2-m_q\zeta\left[m_q^2(9 v^2_3-2v_4)\zeta^2\right.\nonumber \\ &\left.+\mu^2_1\right]z^3\log(z)
+\dfrac{\braket{\bar{q}q}}{\zeta}z^3+\mathcal{O}(z^4),
\end{align}
where $\zeta=\sqrt{3}/(2\pi)$ is the suitable normalization factor as pointed out by \cite{Cherman:2008eh}. Thus, the chiral condensate $\sigma=\braket{\bar{q}q}$ can be extracted by solving corresponding equation of motion, when the quark mass $m_q$ is specified.

\begin{figure}[htbp]
\centering
\vspace{0.5cm}
\subfigure[The temperature dependence of chiral condensate $\sigma$. Different color refers to different phase. The green vertical dashed lines mark the domain where multiple phases coexist. The red vertical line marks the transition temperature $T_c=175.4\text{MeV}$.]{\label{fig:1st PT chiral}\includegraphics[scale=0.23]{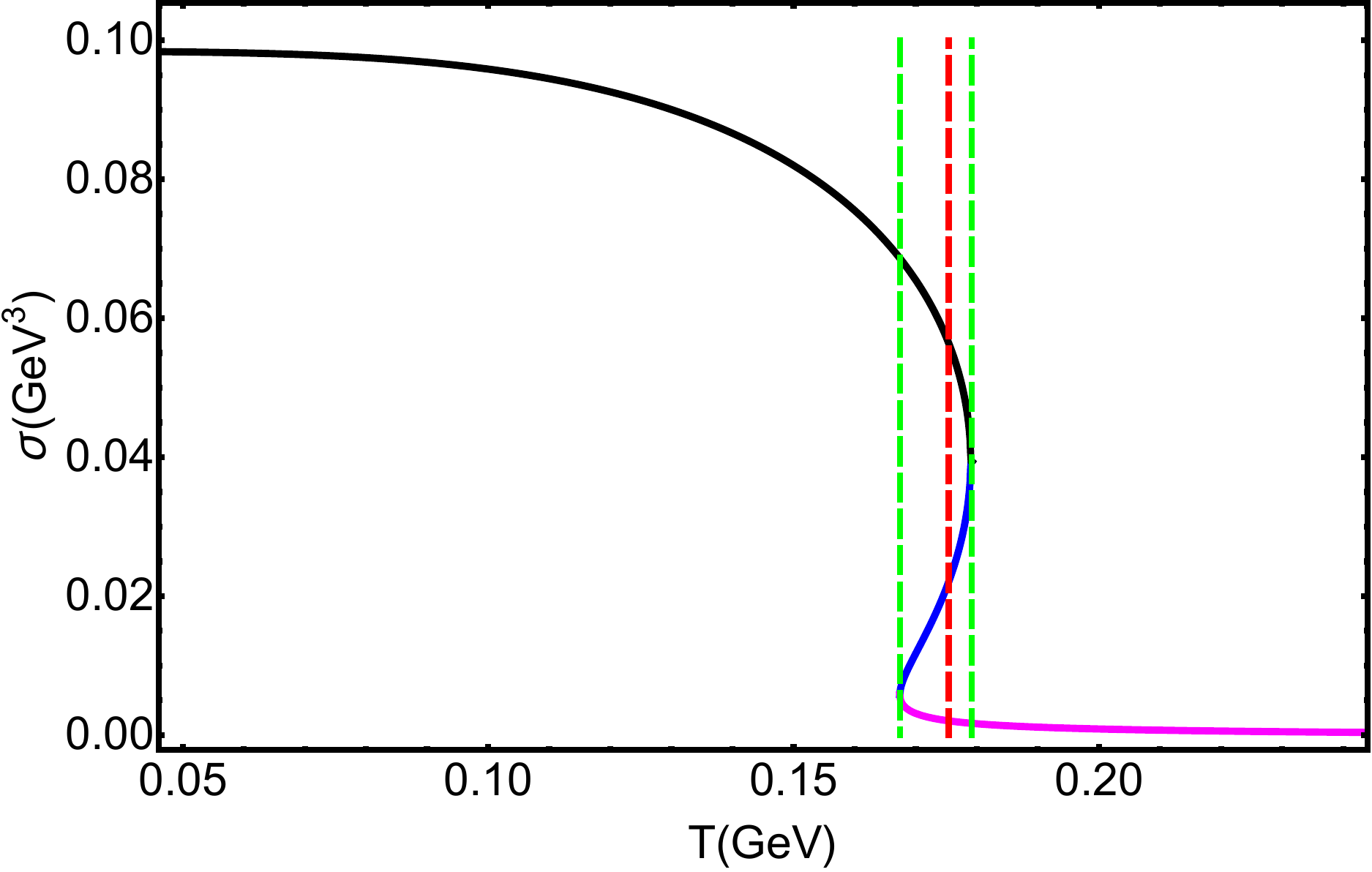}}
\subfigure[The calculated free energy densities with quark mass $m_q=7\text{MeV}$. The color of each curve corresponds exactly to the solution of the same color in the above phase diagram.]{\label{fig:1st PT free energy}\includegraphics[scale=0.23]{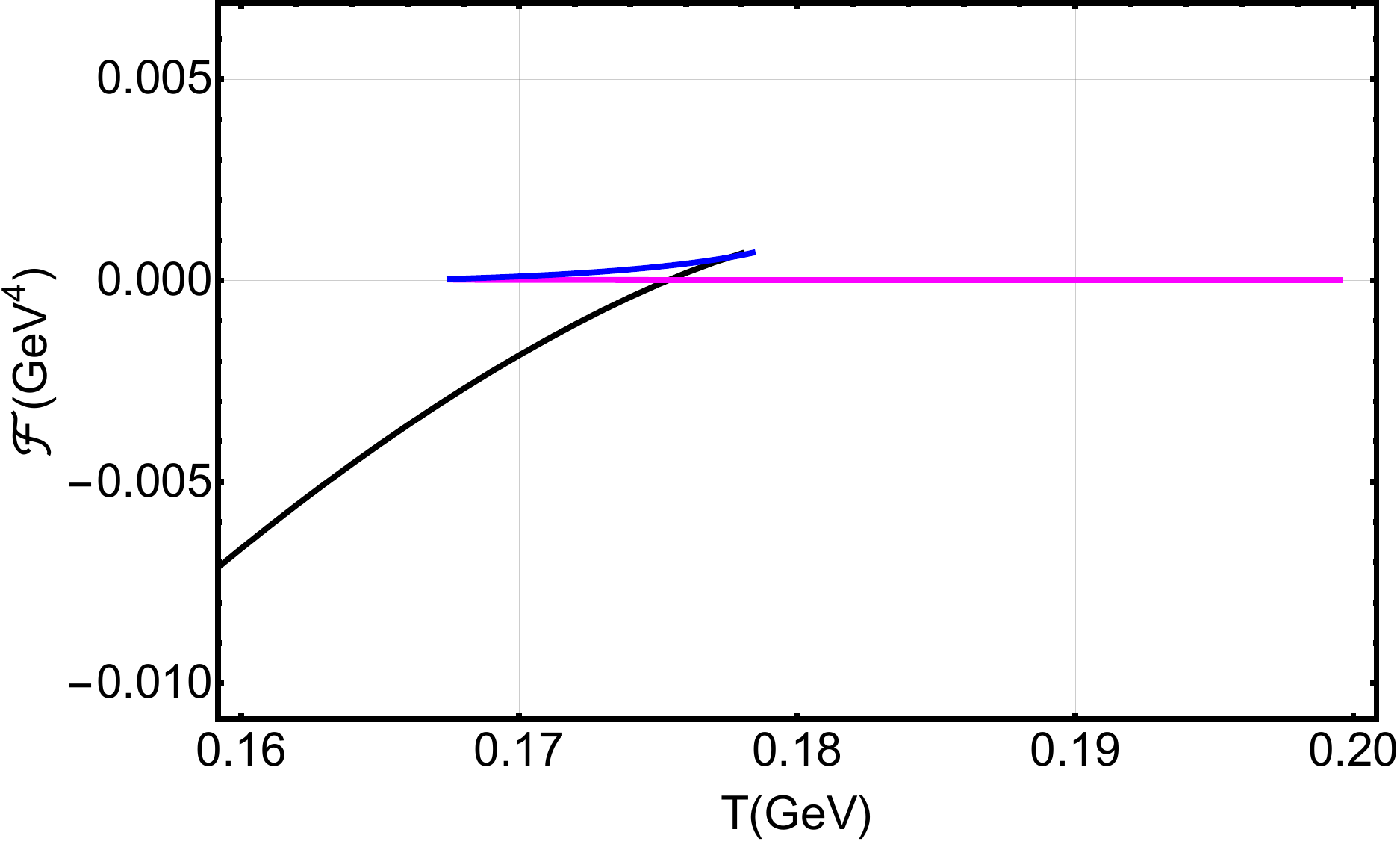}}
\caption{The typical first order chiral phase transition behavior with quark mass $m_q=7\text{MeV}$. These results are the same as the previous studies \cite{Li:2016smq}.}
\label{fig:1st PT}
\end{figure}

In Fig. \ref{fig:1st PT chiral}, we show the temperature dependence of chiral condensate $\sigma$ with the quark mass $m_q=7\text{MeV}$. In this case, the system shows the clear behavior of first order chiral phase transition, with regions containing multiple phases marked by green vertical dashed lines. By using holographic prescription, the free energy density can be obtained by on-shell bulk action. With a finite quark mass, the on-shell action is divergent at $z=0$. We need to do holographic renormalization. After subtracting possible divergent terms by suitable holographic renormalization operation, the calculated results of free energy densities of the system are shown in Fig. \ref{fig:1st PT free energy}. The color of each curve corresponds exactly to the solution of the same color in the phase diagram Fig. \ref{fig:1st PT chiral}. The transition temperature for this first order phase transition can be identified to be $T_c=175.4\text{MeV}$ from points where the free energy densities are equal. The two green vertical dashed lines mark the domain where multiple phases coexist. The concrete temperature values corresponding to these green dashed lines are $T=167.4\text{MeV}$ and $T=179.2\text{MeV}$, respectively.

By varying the quark mass, the chiral phase transition evolves from first-order to second-order, and finally becomes a crossover. This phenomenon is consistent with typical prediction made by Columbia plot \cite{Fischer:2018sdj}. Therefore, employing this model enables us to further investigate the associated dynamical properties of systems exhibiting distinct chiral phase transition behaviors. These potential dynamical properties constitute the primary focus of this work and will be further discussed in subsequent sections.

\section{Linear perturbation and quasi-normal modes\label{sec:linear perturbation}}

Building on the background with varied phase transition behaviors discussed in previous sections, we now examine linear perturbations on the given background.

In this work, we only investigate the scalar mode. Therefore, we decompose the $\chi$ field as follows:
\begin{equation}
    \label{eq:field decomposition}
    \chi(z,t,\bm{x})=\chi(z)+\sigma(z,t,\bm{x}),
\end{equation}
where $\chi(z)$ is just the background field satisfying equation of motion \eqref{eq:background EOM} and $\sigma(z,t,\bm{x})$ is the perturbation field.

To derive the linearized perturbation equations, we retain terms up to second order in the perturbation field $\sigma$ within the 5D bulk action, which can be written as
\begin{align}
\label{eq:perturbation action}
S=&-\int d^5x\sqrt{-g}e^{-\Phi}\left[\dfrac{1}{2} g^{MN}\partial_{M}\sigma\partial_N\sigma+g^{zz}\partial_z\chi\partial_z\sigma
\right.\nonumber\\
&\left.+\left(-3\chi+3v_3\chi^2+4v_4\chi^3\right)\sigma+\left(-\frac{3}{2}+3v_3\chi+6v_4\chi^2\right)\sigma^2 \right].
\end{align}
Given that the background field $\chi(z)$ must satisfy its own on-shell condition, the corresponding linearized perturbation equation take the following form:
\begin{align}
    \label{eq:linearized equation}
    0=&\sigma^{\prime\prime}_k-\left[\dfrac{3}{z}-\dfrac{f^{{\prime}}(z)}{f(z)}+\Phi^{\prime}(z)\right]\sigma^{\prime}_k+\left(\dfrac{\omega^2}{f(z)^2}
    -\dfrac{\bm{k}^2}{f(z)}\right)\sigma_k\nonumber\\
    &-\dfrac{1}{z^2f(z)} (-3+6v_3 \chi+12v_4\chi^2)\sigma_k.
\end{align}
Here, owing to the linear nature of the perturbation equation, modes with different momenta decouple from each other. Consequently, the momentum-space representation of the perturbation field can be safely substituted for its corresponding coordinate-space field without introducing any inconsistencies.

The corresponding on-shell action will reduce to the boundary term
\begin{align}
    \label{eq:perturbation on-shell}
    S_{\text{boundary}}\sim& -\dfrac{1}{2}\lim_{z\rightarrow0}\int \dfrac{d^4k}{(2\pi)^4} \left[\sqrt{-g}e^{-\Phi} g^{zz} \sigma_k(z)\partial_z\sigma_k(z)\right] \nonumber\\
    &+\text{contact terms}.
\end{align}
On the other hand, the solution obeying the incoming boundary condition at the horizon can be written as
\begin{equation}
    \label{eq:perturbation boundary}
    \sigma_k(z)=s_1(k)z(1+\cdots)+s_3(k)z^3(1+\cdots).
\end{equation}
In holographic duality, the perturbation field $\sigma$ couples to some particular operator $\mathcal{O}$ of the dual field theory at the boundary. Then, by applying the Lorentzian AdS/CFT prescription developed in \cite{Son:2002sd} to the on-shell action \eqref{eq:perturbation on-shell}, the retarded correlator can be given by 
\begin{equation}
    \label{eq:retarded function}
    G^R(k)\equiv\braket{\mathcal{O}(k)\mathcal{O}(-k)}_R\sim \dfrac{s_3(k)}{s_1(k)}+\text{contact terms}.
\end{equation}
Consequently, the poles of retarded correlator correspond to zeros of the function $s_1(k)$. It should be emphasized that when solving the perturbation equation \eqref{eq:linearized equation}, incoming boundary condition must be imposed at the horizon. This is equivalent to basic definition of quasi-normal modes. In other words, condition $s_1(k)=0$ actually defines the spectrum of quasi-normal modes for scalar perturbations. This statement was first pointed also by \cite{Son:2002sd}. The real and imaginary parts of the quasi-normal frequencies can be mapped to the pole masses and the thermal widths of particles excited from the hot medium \cite{Miranda:2009uw,Grigoryan:2010pj,Mamani:2013ssa,Braga:2019yeh,Cao:2021tcr}. 

For the convenience of forthcoming analysis, we also define the corresponding spectral function as
\begin{equation}
    \label{eq:spectral definition}
    \rho(k)=-2\text{Im} G^R(k),
\end{equation}
which is a commonly used definition.

In summary, we will use the framework outlined above, calculate the quasi-normal spectrum and retarded correlator (more precisely, spectral function) with given background configurations.

\section{Dynamical instability in the spinodal region\label{sec:dynamical instability}}

\begin{figure}[htbp]
\centering
\vspace{0.2cm}
\includegraphics[scale=0.4]{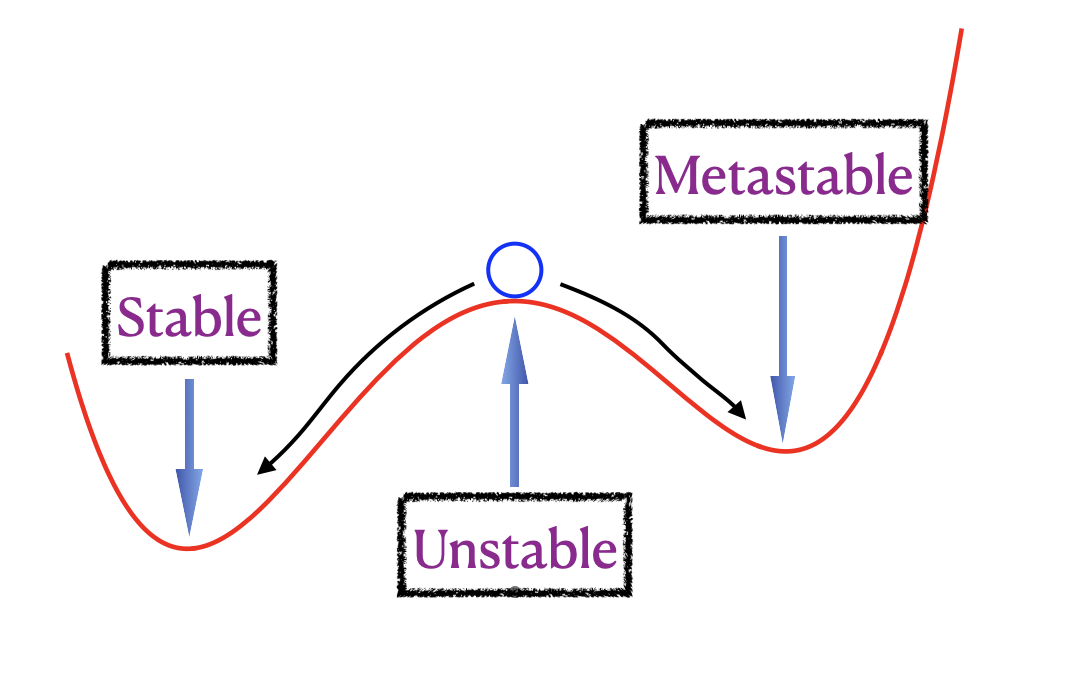}
\caption{The schematic picture of a typical first-order phase transition in the spinodal region. The red curve represents the usual thermodynamic potential of system of interest. The thermodynamically unstable solution refers to the local maximum of this potential.}
\label{fig:basic picture}
\end{figure}

Inspired by Gubser and Mitra \cite{Gubser:2000ec,Gubser:2000mm}, systems with thermodynamic instability may induce possible dynamic instability. Based on the previous introduction, we can construct a system exhibiting first-order chiral phase transition behavior using the nonlinear soft-wall model, its phase diagram is already shown in Fig. \ref{fig:1st PT}. Note that a hallmark of first-order phase transitions is the existence of thermodynamically unstable solution within their spinodal region. The basic schematic picture is shown in Fig. \ref{fig:basic picture}. Thus, we can use current model to study possible dynamical instability induced by thermodynamic instability directly. In this section, we will discuss such dynamical instability in the spinodal region, either at linear perturbation level or at fully nonlinear evolution level.

\subsection{Linear dynamical instability from quasi-normal modes}
\label{subsec:linear QNM analysis}

At linear level, the dynamical properties of a certain system can be well described by its quasi-normal modes. In general, during the initial transient regime, the non-equilibrium evolution of a system exhibits significant sensitivity to initial conditions and thus fails to reveal its intrinsic dynamic properties. Consequently, we focus on the long-time behavior, and equivalently, the lowest quasi-normal mode in the whole spectrum. Physically, the inverse of the frequency of such lowest quasi-normal mode can be interpreted as the upper bound of thermalization time of given system.

Typically, the time dependent behavior of certain mode can be approximately described by $e^{-i\omega t}$, especially in the long-time regime. Remember that the frequencies of quasi-normal modes are complex numbers. Thus, in the analysis based on quasi-normal modes, if the frequency of a quasi-normal mode has a positive imaginary part, then the corresponding mode will exhibit exponential temporal growth $\sim e^{|\text{Im}[\omega]|t}$, inducing system instability that culminates in fragmentation. On the contrary, those perturbation modes with negative imaginary part will decay exponentially and lead the system relax back to the given background solution. Hence, we adopt the presence of quasi-normal modes with positive imaginary parts as the criterion for determining system dynamical stability under linear perturbations. 

\begin{figure}[htbp]
\centering
\vspace{0.3cm}
\includegraphics[scale=0.27]{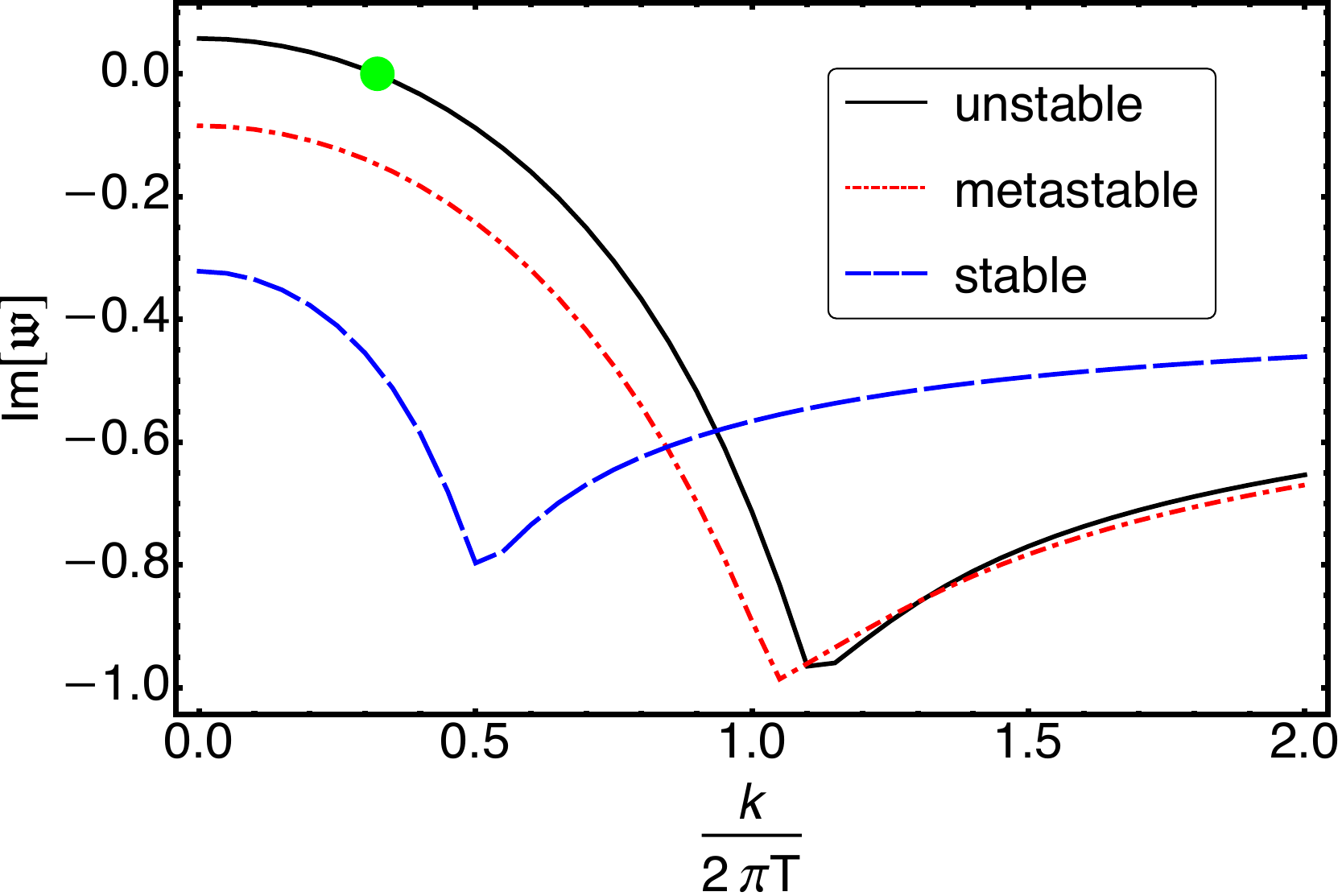}
\caption{The imaginary part of normalized frequency $\mathfrak{w}=\omega/2\pi T$ of lowest QNM as a function of dimensionless spatial momentum $k/2\pi T$ with respect to three different background solutions at $T=173.3\text{MeV}$. The large green dot refers to the position corresponding to the critical momentum $k_c=0.32(2\pi T)=348.4\text{MeV}$.}
\label{fig:1733 dynamical instability}
\end{figure}

In Fig. \ref{fig:1733 dynamical instability}, we show the spatial momentum dependence of the imaginary part of lowest quasi-normal modes at fixed temperature $T=173.3\text{MeV}$. All the parameters are normalized to be dimensionless. Compared to Fig. \ref{fig:1st PT}, this temperature lies in the spinodal region, and consequently, we have three different background solutions corresponding to different static thermodynamic properties. Clearly, dynamic instability can potentially emerge exclusively from perturbations applied to thermodynamically unstable backgrounds, which is denoted by black line in Fig. \ref{fig:1733 dynamical instability}. For comparison, for those thermodynamically stable backgrounds, all corresponding quasi-normal modes have frequencies with negative imaginary parts, which physically correspond to dynamically stable perturbations. This result agrees with our expectation based on Gubser-Mitra conjecture.

Interestingly, possible dynamic instabilities exist only within finite momentum-space regions. The large green dot in the Fig. \ref{fig:1733 dynamical instability} marks the position of critical momentum $k_c$, which is $348.4\text{MeV}$ in this case. Note that physically, the critical momentum corresponds to a critical length scale $l_c\sim 1/k_c$. Thus, the existence of a critical momentum implies that only perturbations with sufficiently large spatial scales ($l>l_c$) can induce dynamic instability to the entire system. As a result, given that small-scale perturbations remain dynamically stable, thermodynamically unstable state retains a degree of stability in some sense. The observation described above shares similar qualitative characteristic with previous studies on Gregory-Laflamme instability \cite{Gregory:1993vy,Gregory:1994bj} and spinodal instability of holographic superfluid \cite{Zhao:2023ffs,Zhao:2022jvs}. But it should be emphasized that in the previous studies mentioned above, all perturbation modes were exclusively hydrodynamic modes, satisfying $\lim_{k\rightarrow 0}\omega(k)=0$. In our case, it is obvious that the perturbation modes are all non-hydrodynamic modes. This phenomenon may be interpreted as the consequence of the infinite 't Hooft coupling \cite{Casalderrey-Solana:2018rle}.
 
\begin{figure}[htbp]
\centering
\vspace{0.3cm}
\includegraphics[scale=0.27]{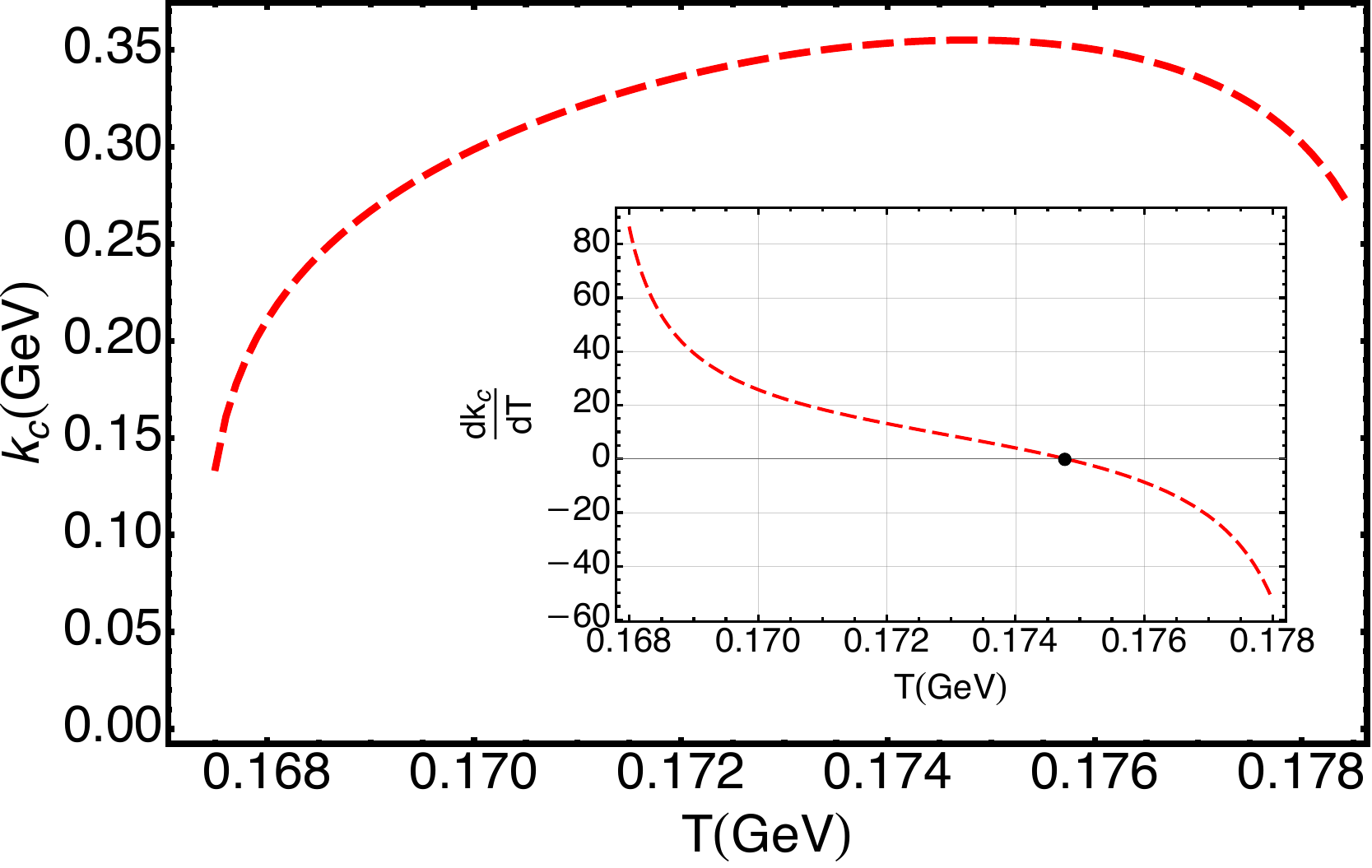}
\caption{The temperature dependence of critical momentum $k_c$. The insert shows the derivative of critical momentum with respect to temperature. The black dot in the insert marks the maximum of critical momentum, with temperature $T_{\text{Max}}=174.8\text{MeV}$.}
\label{fig:critical momentum}
\end{figure}

Besides, the critical momentum also shows the non-trivial temperature dependence. In Fig. \ref{fig:critical momentum}, we directly investigate such dependence. Systems with larger critical momentum exhibit greater susceptibility to induced dynamic instability. Consequently, critical momentum $k_c$ may serve as a quantitative measure of the degree of dynamic instability in a given system. In the case studied in this section, the maximum of $k_c$ appears at $T=174.8\text{MeV}$, which is rather close to the first-order chiral phase transition temperature $T_c=175.4\text{MeV}$. Owing to this, we conclude that the region near the critical temperature of first-order phase transition can be related to the region with maximal dynamical instability. This observation implies that some dynamical information of the given system may be encoded in some thermodynamic quantities.

\subsection{Dynamical instability from nonlinear evolution}
\label{subsec:nonlinear evolution}

Although the analysis based on quasi-normal modes does provide substantial insights into system dynamical properties, it loses all the effects coming from full non-linear coupling. In this section, we briefly analyze the dynamical instability by using full non-linear time dependent evolution. Our purpose is to provide further evidence for the existence of dynamic instability discussed in preceding section.

In general, the bulk geometry with embedding black hole has singularity at the black hole horizon. To avoid such difficulty, suitable coordinate system should be chosen. In this work, we transform to the Eddington-Finkelstein (EF) coordinates, and the corresponding metric can be written as 
\begin{equation}
    \label{eq:EF metric}
    ds^2=\dfrac{1}{z^2}\left[-f(z)dt^2-2dtdz+dx^idx^i\right].
\end{equation}
Here, we use $t$ coordinate as EF time without introducing any ambiguity. The typical advantage is that EF time in bulk reduces to physical time in boundary and remains non-singular at the horizon.

Empirical insights from linear perturbation theory indicate that large-scale perturbations exhibit heightened propensity to trigger dynamical instability. Thus, in this section, we focus on the perturbations with spatial momentum $\bm{k}=0$, which is equivalent to spatially homogeneous perturbations. 

\begin{figure}[htbp]
\centering
\vspace{0.3cm}
\includegraphics[scale=0.27]{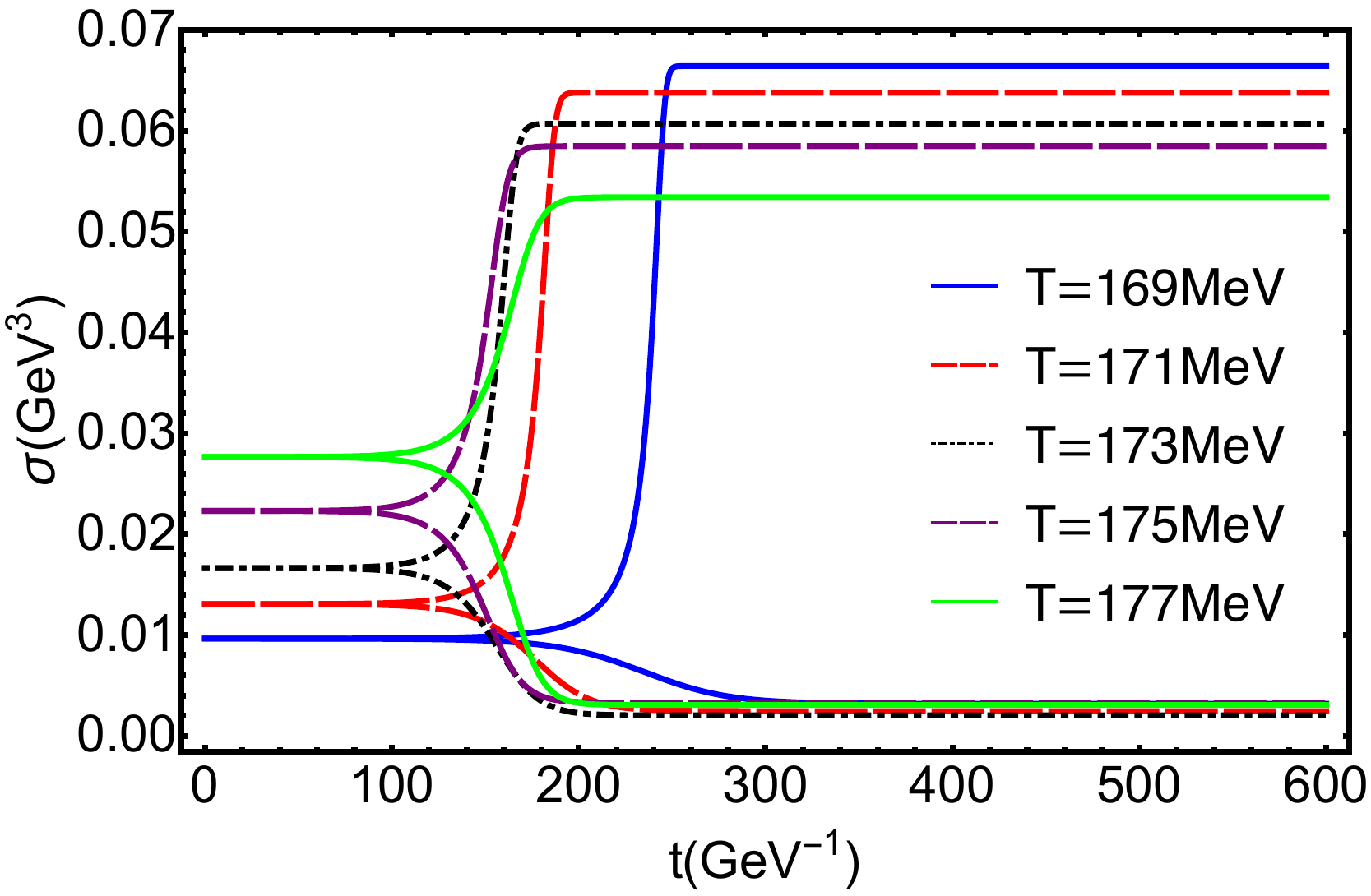}
\caption{The non-linear evolution of chiral condensate $\sigma=\braket{\bar{q}q}$ with different temperature. All the background solutions are chosen to be thermodynamically unstable. }
\label{fig:nonlinear evolution}
\end{figure}

In Fig. \ref{fig:nonlinear evolution}, we show time evolution of chiral condensate at full non-linear level. These temperatures are chosen to be in the spinodal region of first-order phase transition. The background solution corresponding to each curve is chosen to be the thermodynamically unstable one. We set the typical magnitude of the initial perturbations to be of order $\sim\mathcal{O}(10^{-9})$, which is a rather small value. Even under such tiny perturbations, thermodynamically unstable systems manifest unambiguous phase separation after evolving over sufficient timescales. This is the obvious evidence that the modes corresponding to dynamical instability dominate the long-time evolution behaviors of given systems. In this way, the specific time when the phase separation begins can be treated as the measure of dynamical instability. Consistent with the observation from linear perturbation calculation, systems exhibit progressively earlier onset of phase separation driven by dynamic instability when approaching the first-order transition temperature $T_c$.

In summary, either at linear level or at full non-linear level, the background with thermodynamical instability do induce dynamical instability. The closer a system approaches the first-order phase transition temperature, the more readily dynamic instability can be triggered. 

\section{Diffusion-to-sound transition\label{sec:diffusion-to-sound}}

Apart from evidence signaling dynamic instability, some more interesting phenomenon can also be observed in Fig. \ref{fig:1733 dynamical instability}. This phenomenon will constitute the primary focus of in-depth examination in this section. 

\subsection{Basic characteristics of the transition\label{subsec:transition}}

\begin{figure}[htbp]
\centering
\vspace{0.3cm}
\includegraphics[scale=0.27]{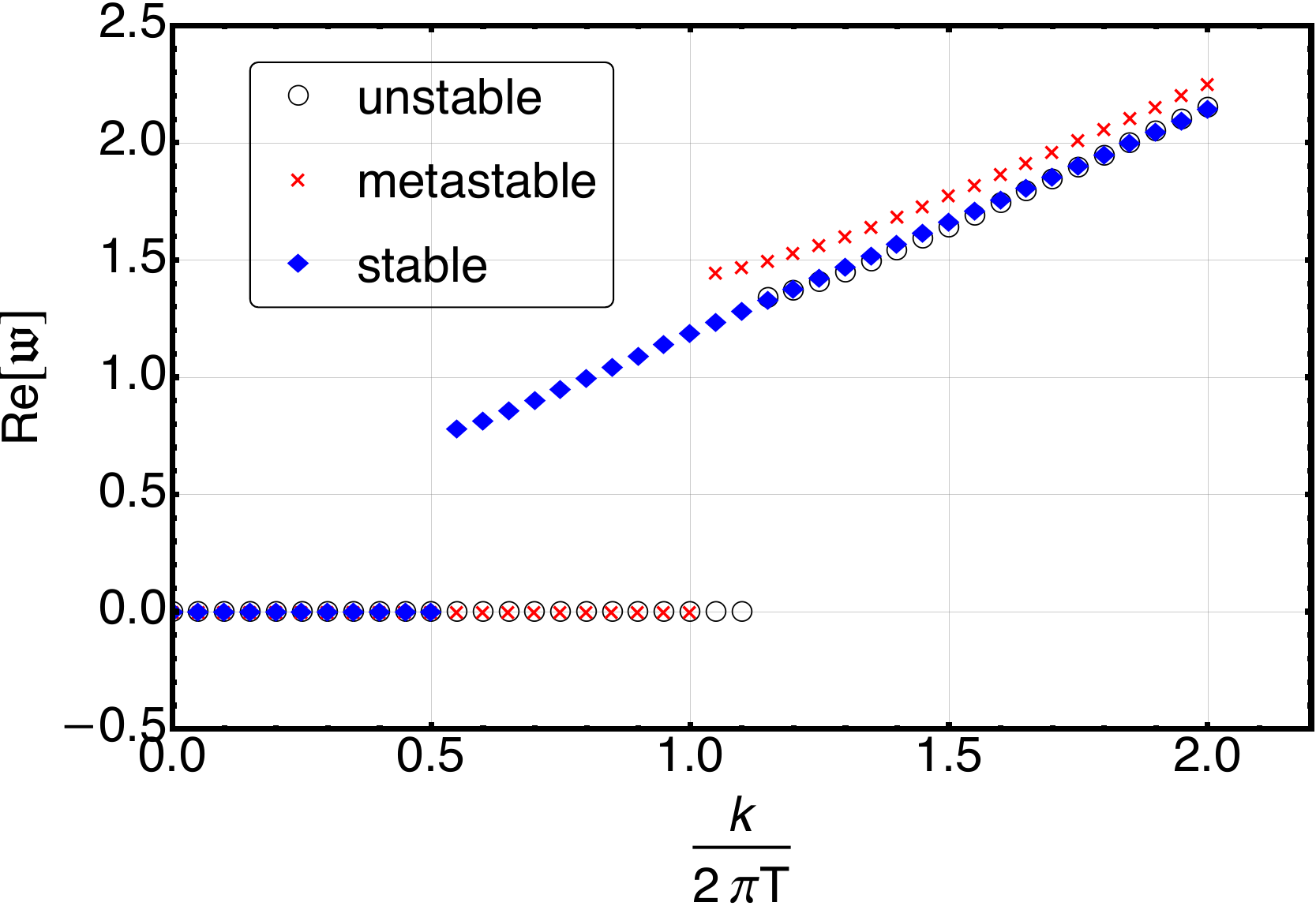}
\caption{The real part of normalized frequency $\mathfrak{w}=\omega/2\pi T$ as a function of dimensionless spatial momentum $k/2\pi T$ for three different solutions at $T=173.3\text{MeV}$.}
\label{fig:1733QNMReal}
\end{figure}

Note that, in Fig. \ref{fig:1733 dynamical instability}, the monotonicity of the imaginary part of lowest quasi-normal mode frequencies with respect to spatial momentum exhibits significant alteration. To uncover the underlying physics hidden behind this phenomenon, in Fig. \ref{fig:1733QNMReal}, we show the spatial momentum dependence of the real part of lowest quasi-normal modes, which are the same modes already shown in Fig. \ref{fig:1733 dynamical instability}. 

Compare Fig. \ref{fig:1733 dynamical instability} with Fig. \ref{fig:1733QNMReal}, at the transition point where monotonicity of the frequency's imaginary part changes, the underlying physical process involves mode transition from diffusive mode (with vanishing real part) to propagating mode (with non-zero real part). This means that the property of mode which is dominant in the long-time evolution region has significantly changed. To illustrate, within the long-time evolution regime, perturbations at large spatial scales (corresponding to small momentum regions) can only approach the final state via diffusive processes, whereas at small spatial scales (corresponding to high momentum regions), they can propagate like usual sound mode.  

It should be emphasized that in Fig. \ref{fig:1733QNMReal}, at identical temperature, perturbation modes on three distinct thermodynamic background solutions exhibit similar diffusion-to-sound transition behavior. This suggests that this typical transition behaviors are not related to the nature of first-order chiral phase transition. Indeed, we check the situation with larger quark masses, which correspond to usual crossover behaviors, and also observe this particular diffusion-to-sound transition. To simplify, we will investigate the case with the background solution exhibiting crossover behavior and set quark mass $m_q=40\text{MeV}$ in the subsequent discussion. After further detailed investigation, although the aforementioned phenomenon shows no apparent correlation with the specific order of the chiral phase transition, its manifestation is fundamentally linked to chiral symmetry breaking and its restoration.   

\begin{figure}[htbp]
\centering
\vspace{0.3cm}
\includegraphics[scale=0.27]{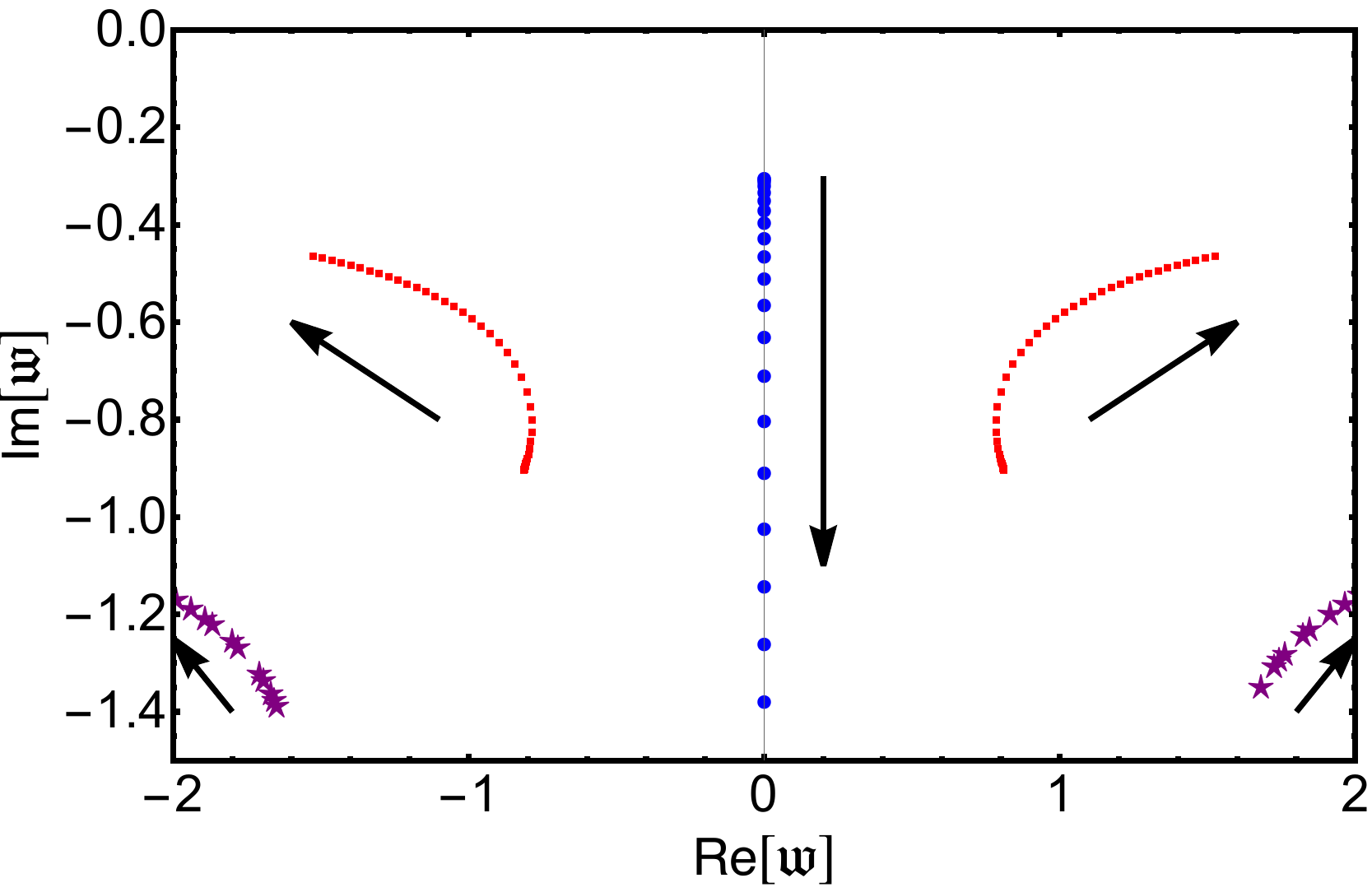}
\caption{Evolution of the lowest five quasi-normal modes of scalar perturbations in the complex dimensionless frequency $\mathfrak{w}=\omega/2\pi T$ plane. The black arrows show the directions corresponding to the spatial momentum increasing.}
\label{fig:240Complex}
\end{figure}

The information obtained from the lowest QNM only is not complete. One may be misled to conclude that the diffusive modes no longer exist in the large spatial momentum region after the manifestation of diffusion-to-sound transition by observing Fig. \ref{fig:1733QNMReal}. This is not exactly the case. To resolve the aforementioned ambiguity, we further study the evolution of the lowest five quasi-normal modes in the complex plane as functions of spatial momentum. In Fig. \ref{fig:240Complex}, we choose the background temperature to be $240\text{MeV}$ and plot the desired quasi-normal modes in the whole complex plane. The rationale for selecting a relative high background temperature stems from the predominance of propagating modes over diffusive modes within low-temperature regimes. Consequently, to investigate the interplay between diffusive and propagating modes, we strategically designate $240 \text{MeV}$ as the representative background temperature.

In Fig. \ref{fig:240Complex}, one direct observation is that the diffusive mode does not simply disappear in the large momentum region. Such mode always manifests itself in the quasi-normal spectrum when the chiral system is in high-temperature regime. The fundamental distinction underlying diffusion-to-sound transition lies in the starkly contrasting dependence on spatial momentum exhibited by the imaginary parts of diffusive versus propagating modes. Namely, as spatial momentum increases, the absolute value of the imaginary part of diffusive mode increases, and absolute value of the imaginary part of propagating mode, by contrast, decreases. Consequently, when the spatial momentum exceeds a critical value, the absolute value of the imaginary part becomes greater for diffusive modes than for propagating modes. Typically, the mode with negative imaginary part decays exponentially in time $\sim e^{-|\text{Im}[\omega]|t}$. Thus, mode with larger absolute imaginary part experiences faster temporal decay, as a result, such modes ceases to dominate in long-time evolution.  To summarize, diffusive mode undergoes rapid decay in high-momentum regimes, thus allowing propagating modes to be dominant in long-time evolution within these momentum domains. This characteristic is captured by the diffusion-to-sound transition observed in Fig. \ref{fig:1733QNMReal}. 

\subsection{Spectral functions and transport peak\label{subsec:transport peak}}

While our preceding discussion has detailed the fundamental aspects of the diffusion-to-sound transition phenomenon, a thorough exploration of possible underlying physical interpretation has yet to be undertaken. In this section, we calculate corresponding spectral functions and try to give a simple physical picture of such interesting transition.

The basic framework suitable for calculating spectral functions has been already introduced in Sec. \ref{sec:linear perturbation}, specific implementation details of the calculations are not reiterated here. But we should emphasized that the corresponding operator we calculate in this section is the scalar operator $\hat{\mathcal{O}}(x)=\bar{q}q(x)$. 

\begin{figure*}[htbp]
  \centering
  \begin{minipage}{0.45\textwidth}
    \centering
    \includegraphics[scale=0.27]{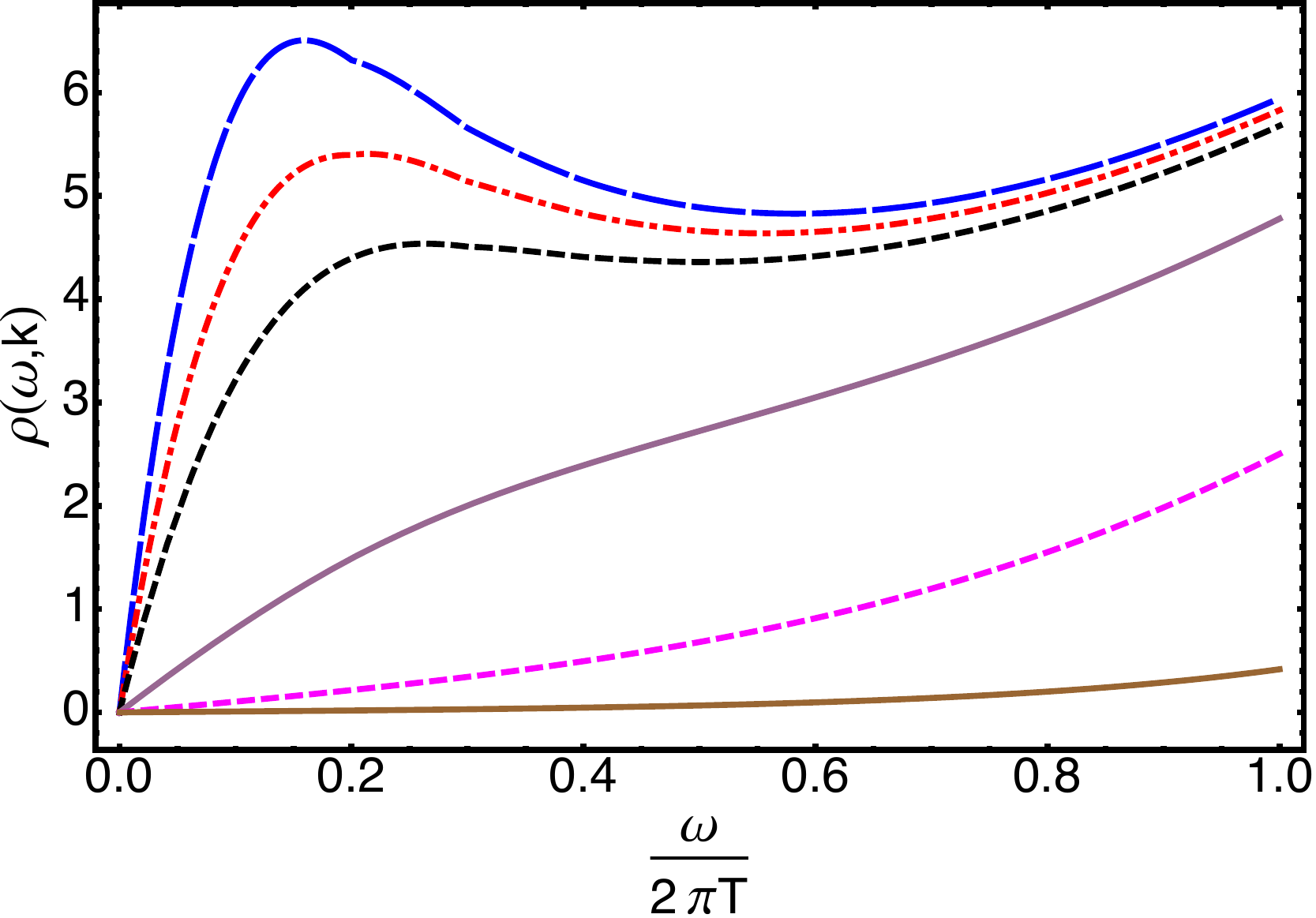}
  \end{minipage}
  \hspace{5mm}
  \begin{minipage}{0.45\textwidth}
    \centering
    \includegraphics[width=\textwidth]{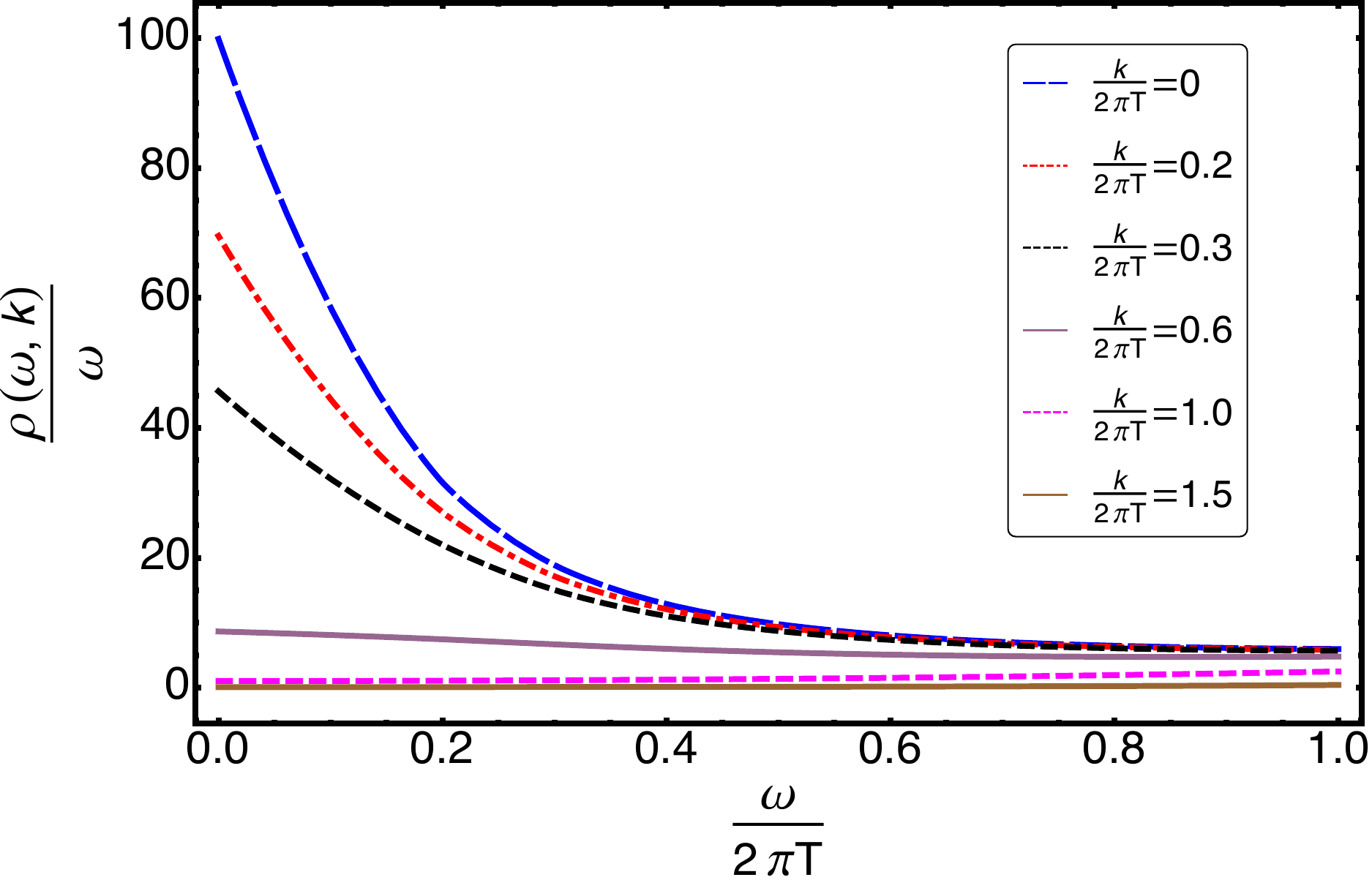}
  \end{minipage}
  \caption{The spectral function of scalar operator $\hat{\mathcal{O}}(x)=\bar{q}q(x)$ as a function of dimensionless frequency $\omega/2\pi T$ for different spatial momentum. The background temperature is chosen to be $200\text{MeV}$. The left panel shows behaviors of the original spectral function $\rho(\omega,k)$ and the right panel shows behaviors of the rescaled spectral function $\rho(\omega,k)/\omega$. The two panels share the same figure legend.}
  \label{fig:200Spectral}
\end{figure*}

We first investigate the influence of spatial momentum on the spectral functions. To do this, we calculate the desired spectral function $\rho(\omega,k)$ as a function of dimensionless frequency $\mathfrak{w}=\omega/2\pi T$ for different spatial momentum. The results are displayed in Fig. \ref{fig:200Spectral}. In the right panel of Fig. \ref{fig:200Spectral}, we rescale the spectral function by a factor $\omega$ in order to elucidate the behavior within the low-frequency region. In this case, a narrow structure known as the transport peak (regarding studies on transport peak, one may consult Refs. \cite{Casalderrey-Solana:2018rle,Petreczky:2005nh,Aarts:2002cc}) rises in the small frequency region. Further, We have moderately expanded the conceptual scope of transport peaks. Typically, the transport peak is defined at zero spatial momentum. In this work, we treat similar peaks appearing in the spectral functions as transport peaks, even though the corresponding spatial momentum $k$ are not zero.      

One direct observation is that such narrow peak only exists in the small spatial momentum region, $0\leq k\leq k_{max}$. This aligns precisely with the earlier observation from quasi-normal mode computations: diffusive modes dominate only within low spatial momentum regions. Clearly, these narrow peaks correspond to such diffusive modes. The correspondence between diffusive modes and transport peak structures in spectral functions has been reported in prior research \cite{Casalderrey-Solana:2018rle}.

\begin{figure}[htbp]
\centering
\vspace{0.3cm}
\includegraphics[scale=0.27]{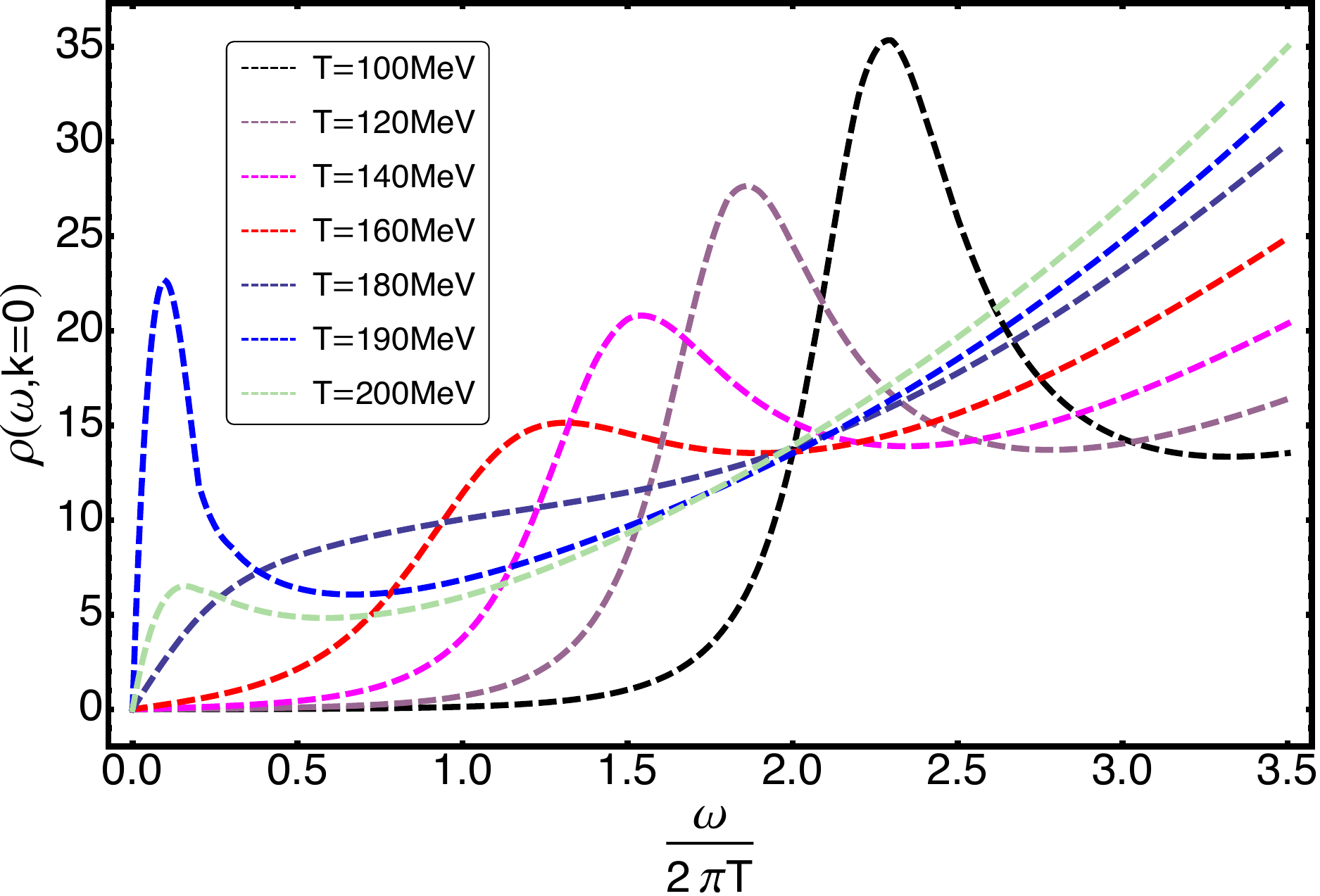}
\caption{The temperature dependence of spectral functions as functions of dimensionless $\omega/2\pi T$ for zero spatial momentum.}
\label{fig:peak temperature}
\end{figure}

Next we examine the influence of temperature on transport peak structure. By investigating the temperature dependence of spectral functions, we aim to establish connections between the emergence of transport peak structures and the breaking/restoration of chiral symmetry. In order to simplify the situation, we restrict our calculation for zero spatial momentum. Our numerical results are displayed in Fig. \ref{fig:peak temperature}.

When temperature is low, the spectral functions exhibit peaks which satisfy clear Breit-Wigner distribution form. These signify the presence of bosonic excitation modes in the system exhibiting a well-defined thermal mass and finite thermal width. Remembering the scalar operator $\hat{\mathcal{O}}(x)=\bar{q}q(x)$ we have used in this work, these excitations can be recognized as scalar $\sigma$ meson excitation. Compared with other mesons in thermal medium (for example, $\pi$ meson \cite{Cao:2022csq}), scalar $\sigma$ meson demonstrates significantly broader thermal widths, consequently exhibiting shorter lifetimes. With increasing temperature, the background chiral condensate $\braket{\bar{q}q}$ is reduced and the broken chiral symmetry tends to restore. Consequently, the thermal width of scalar $\sigma$ meson increases significantly. Because of this, the final fate of thermal $\sigma$ meson is dissociation. In Fig. \ref{fig:peak temperature}, at $T=180\text{MeV}$, the spectral function no longer exhibits discernible excitation peaks corresponding to such scalar mesons. This means that at $T=180\text{MeV}$, the thermal scalar $\sigma$ meson has completely dissociated. Further increasing the temperature reveals intriguing phenomena. In the near zero frequency region, a new peak manifests itself and the height of the peak reaches its maximum value when $T\sim 190\text{MeV}$. This kind of peak is just aforementioned transport peak. This correspondence, where spectral functions exhibit meson excitation peaks at low temperatures but transport peaks at high temperatures, aligns with the temperature-dependent evolution of the lowest quasi-normal mode. 

\begin{figure}[htbp]
\centering
\vspace{0.3cm}
\subfigure[The temperature dependence of real part of lowest QNM frequency $\mathfrak{w}=\omega/2\pi T$.]{\label{fig:lowestQNM Real}\includegraphics[scale=0.25]{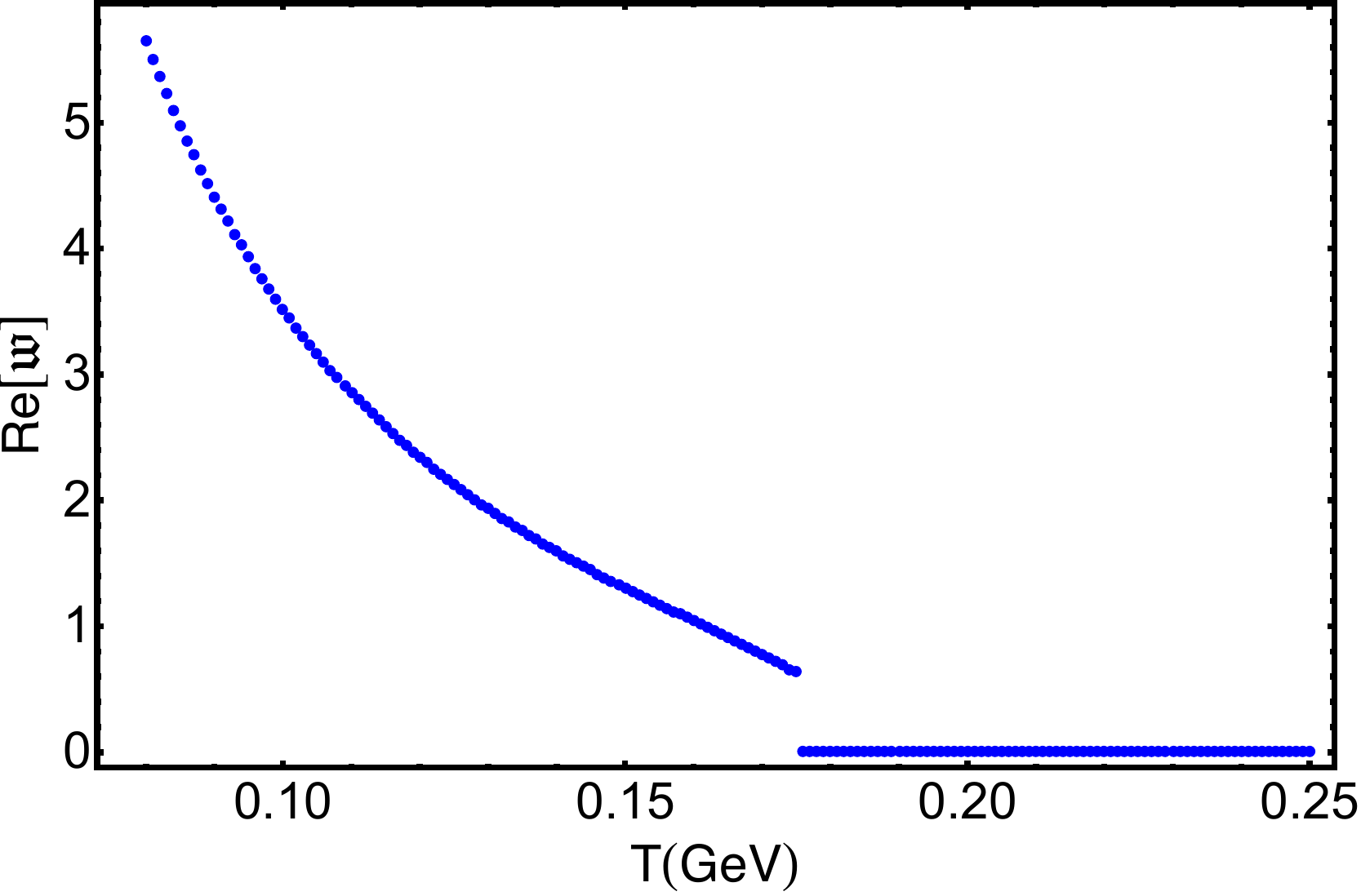}}
\subfigure[The temperature dependence of imaginary part of lowest QNM frequency $\mathfrak{w}=\omega/2\pi T$.]{\label{fig:lowestQNM Imaginary}\includegraphics[scale=0.25]{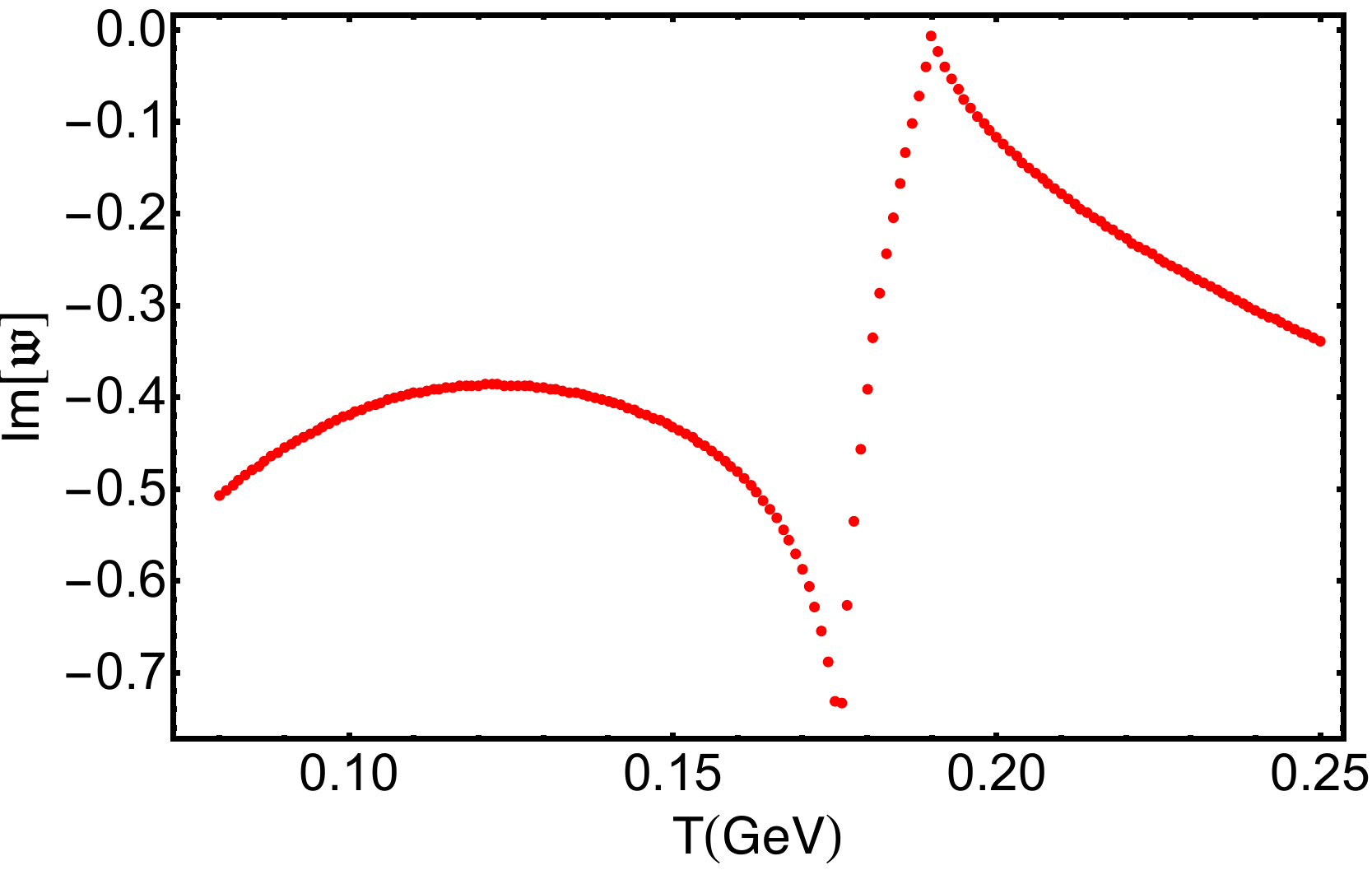}}
\caption{The temperature dependence of lowest QNM frequency $\mathfrak{w}=\omega/2\pi T$. The spatial momentum $k$ is set to be zero.}
\label{fig:lowest QNM}
\end{figure}

\begin{figure}[htbp]
\centering
\vspace{0.3cm}
\includegraphics[scale=0.23]{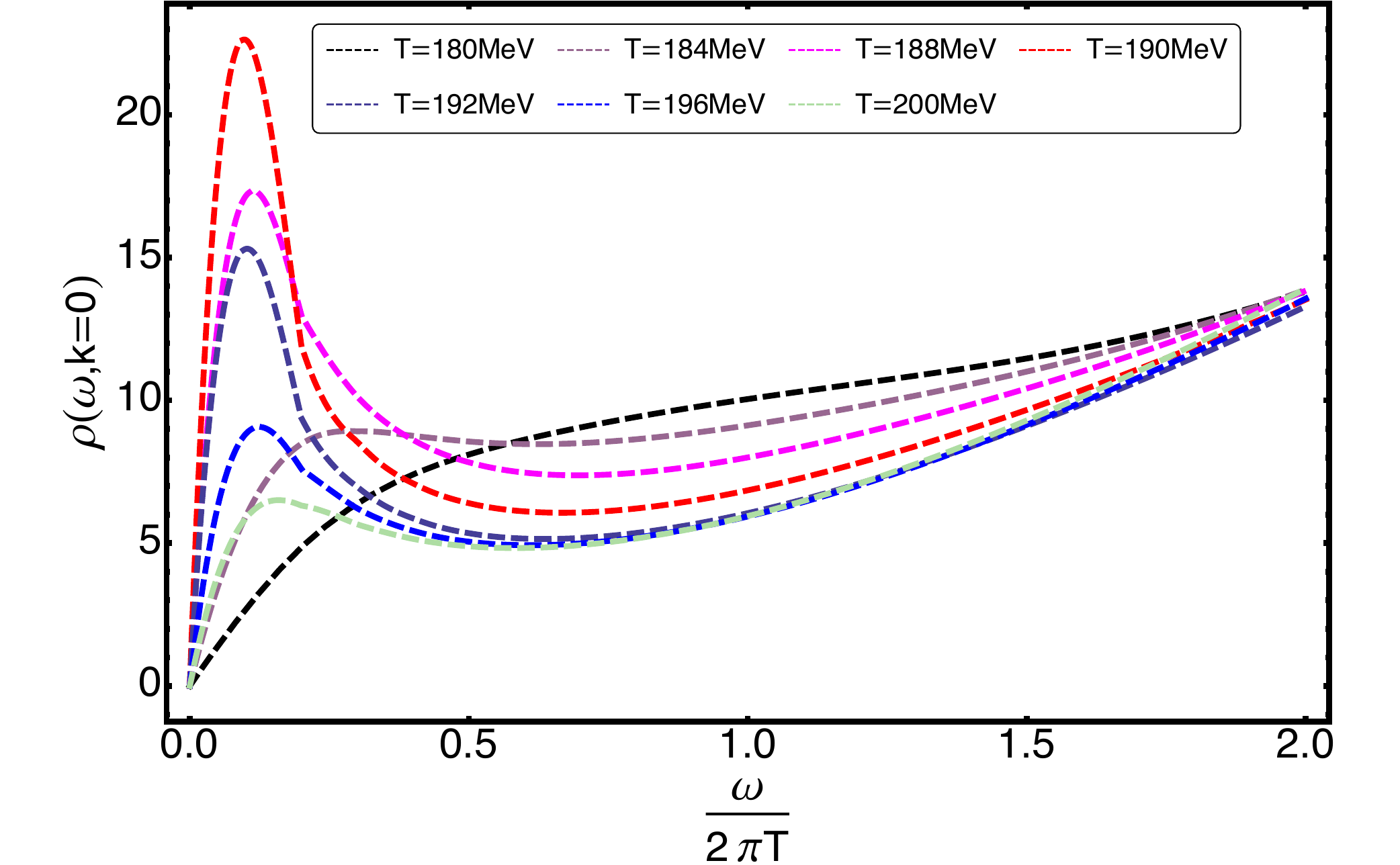}
\caption{The temperature dependence of spectral functions as functions of dimensionless $\omega/2\pi T$ for zero spatial momentum in specific temperature region, $180\text{MeV}\leq T \leq 200\text{MeV}$.}
\label{fig:small region}
\end{figure}

To clarify, we show the temperature dependence of the lowest QNM in Fig. \ref{fig:lowest QNM}. We observe that the so-called diffusion-to-sound transition happens at $T_c\approx 175\text{MeV}$. The diffusive modes emerge in the quasi-normal mode spectrum exclusively above this critical temperature $T_c$. Obviously, this is consistent with the evolution behavior of spectral functions under temperature changing. Interestingly, the imaginary part of lowest QNM increases with temperature in particular temperature region $175\text{MeV}\leq T \leq190\text{MeV}$. In this region, the height of the transport peak also increases. Accordingly, at $T=190\text{MeV}$, the amplitude of transport peak attains its maximum value. Further, the spectral functions in the region $180\text{MeV}\leq T \leq 200\text{MeV}$ are shown in Fig. \ref{fig:small region}. Within this specific temperature range, the temperature dependence of the transport peak amplitude exhibits parallel evolution with the imaginary part of the lowest quasi-normal mode.  

In summary, the emergence of transport peaks in spectral functions occurs subsequent to the complete  dissociation of corresponding scalar $\sigma$ meson excitations, while concurrently, diffusive modes become the governing modes for long-time evolution within the quasi-normal mode spectrum. Naturally, the dissociation of scalar $\sigma$ mesons occurs during partial restoration of chiral symmetry, specifically, when the value of the background chiral condensate falls below a critical threshold, the chiral system ceases to support such mesonic excitations. After the complete dissociation, the basic degrees of freedom of chiral system can be identified as light quarks moving in the thermal medium. Of course, possible gluon contributions should be considered in a thorough investigation, but in this work, we only study contributions from operator constructed by fermionic fields. Further, considering that in prior discussions regarding spatial momentum effects, diffusive modes dominate only for large-scale perturbations. All these fundamental properties indicate that such diffusive modes ultimately emanate from dissipative effects caused by the coupling between non-zero background chiral condensate $\braket{\bar{q}q}$ and the propagating quarks in the medium. From this perspective, the occurrence of the diffusion-to-sound transition should depend solely on the concrete value of the background chiral condensate, thus its observed independence from the specific order of any potential chiral phase transition is justified. From other perspective, in Fig. \ref{fig:small region}, upon further temperature increasing (for instance, $T>190\text{MeV}$), the transport peak begins to diminish. This occurs because further temperature increasing progressively restores chiral symmetry, diminishing the background chiral condensate and thereby reducing the dissipative effects that impede light quark propagation.

\section{Summary and discussion\label{sec:sum}}

In this work, based on a holographic model, we examine the existing of possible dynamical instability in the spinodal region of first-order phase transition. According to the calculated QNM results, the dynamical instability only appears with respect to the background which lacks thermodynamical instability. A defining characteristic of such dynamical instability is that it manifests only within the low-momentum regime, i.e., $0\leq k\leq k_c$. This critical momentum $k_c$ serves as a quantitative measure of the magnitude of dynamical instability. Moreover, critical momentum $k_c$ shows non-trivial temperature dependence. It follows that the region near the first-order phase transition temperature exhibits maximal dynamical instability. These results are consistent with the behavior derived from nonlinear real-time dynamical evolution.

An intriguing additional finding is that the lowest quasi-normal mode undergoes a diffusion-to-sound transition in some temperature region. This means that the dominant mode governing the system's long-time evolution undergoes a transition. As can be seen, diffusive modes only dominate for the perturbations in large spatial scale. When the diffusive mode becomes dominant, the corresponding spectral function exhibits a distinct transport peak. Considering the influence of temperature, we see that the transport peak begins to emerge after complete dissociation of scalar $\sigma$ meson. We argue that this transition phenomenon can be elucidated through the dissipative effects engendered by a nonzero background chiral condensate $\braket{\bar{q}q}$. Based on such interpretation, diffusion-to-sound transition can be related to chiral symmetry breaking and restoration. Besides, it should be emphasized that this transition is independent of the specific order of the chiral phase transition. 

Although in this work we refer to the narrow peak structure emerging in the low-frequency region of the spectral function as the transport peak, its conventional interpretation remains intrinsically linked to specific transport coefficients. For instance, the spectral function $\rho_{xy,xy}(\omega,q)$ of energy-momentum tensor $T_{xy}$ can be related to shear viscosity $\eta$ by the formula (see, e.g., \cite{Casalderrey-Solana:2018rle})
\begin{equation}
    \rho_{xy,xy}(\omega,0)=2\eta \omega +\mathcal{O}(\omega^3).
\end{equation}
Unfortunately, such a relation has yet to be established in this work. Generally, the dynamics of heavy quarks can be adequately described through classical diffusive processes because of large masses, whereas light quarks exhibit highly relativistic motion that cannot be typically captured by classical approximations. In this work, we investigate light quark system where the manifestation of a diffusive mode in the quasi-normal mode spectrum emerges as an entirely unexpected phenomenon. Whether a diffusion coefficient analogous to that for heavy quark system can be meaningfully introduced for light quark matter under such circumstances remains an open question. We aim to resolve through subsequent investigations.

\begin{acknowledgments}

We thank Yi Lu for his helpful discussion. This work is supported by the National Natural Science Foundation of China (NSFC) Grant Nos: 12305136, 12275108, 12235016, 12221005, 12247107, 12175007, the start-up funding of Hangzhou Normal University under Grant No. 4245C50223204075, and the Fundamental Research Funds for the Central Universities. 

\end{acknowledgments}


\bibliography{dynamical}

\end{document}